\documentclass [12pt]{article}
\usepackage{amsfonts,amsmath,amssymb}
\usepackage{graphicx}

\newfont{\twelvemsb}{msbm10 scaled\magstep1}
\newfont{\eightmsb}{msbm8}
\newfam\msbfam
\textfont\msbfam=\twelvemsb \scriptfont\msbfam=\eightmsb
\catcode`\@=11
\def\Bbb{\ifmmode\let\next\Bbb@\else
\def\next{\errmessage{Use \string\Bbb\space only in math mode}}\fi\next}
\def\Bbb@#1{{\fam\msbfam{{#1}}}}

\newcommand{\be}{\begin{equation}}
\newcommand{\ee}{\end{equation}}
\newcommand{\ba}{\begin{eqnarray}}
\newcommand{\ea}{\end{eqnarray}}

\begin{document}

\sloppy
\renewcommand{\thefootnote}{\fnsymbol{footnote}}
\newpage
\setcounter{page}{1} \vspace{0.7cm}
\vspace*{1cm}
\begin{center}
{\bf Beyond cusp anomalous dimension from integrability in SYM$_4$}\\
\vspace{1.8cm} {\large Davide Fioravanti $^a$, Paolo Grinza $^b$ and
Marco Rossi $^c$
\footnote{E-mail: fioravanti@bo.infn.it, pgrinza.grinza@usc.es, rossi@cs.infn.it}}\\
\vspace{.5cm} $^a$ {\em Sezione INFN di Bologna, Dipartimento di Fisica, Universit\`a di Bologna, \\
Via Irnerio 46, Bologna, Italy} \\
\vspace{.3cm} $^b${\em Departamento de Fisica de Particulas,
Universidad de Santiago de Compostela, 15782 Santiago de
Compostela, Spain} \\
\vspace{.3cm} $^c${\em Dipartimento di Fisica dell'Universit\`a
della Calabria and INFN, Gruppo collegato di Cosenza, I-87036
Arcavacata di Rende, Cosenza, Italy} \\
\end{center}
\renewcommand{\thefootnote}{\arabic{footnote}}
\setcounter{footnote}{0}

\begin{abstract}
 We study the first sub-leading correction $O((\ln s)^0)$ to the cusp anomalous dimension  in the high spin expansion of finite twist operators in  ${\cal N}=4$ SYM theory. This term is still governed by a  linear integral equation which we study in the weak and strong coupling regimes. In the strong coupling regime we find agreement with the string theory computations.
\end{abstract}

\vspace{2cm}
\noindent {\it Contribution to the proceedings of the workshop Diffraction 2010, Otranto, 10th-15th September. \\
Talk given by M.Rossi.} \\

\vspace{3cm}

\newpage

\section{Aims, motivations and tools}

Our aim is to compute anomalous dimensions $\gamma $ of linear combinations of twist $L$ scalar operators in planar ${\cal N}=4$ SYM,
\begin{equation}
\mathcal{O}(x)=\sum _{k_1+...+k_L=s}c_{k_1,...,k_L}{\mbox {Tr}} \Bigl ({\cal D}_+^{k_1} \phi (x) ... {\cal D}_+^{k_L} \phi (x)\Bigr ) \, , \label {oper}
\end{equation}
which diagonalise the dilatation operator
\begin{equation}
\hat D\, {\mathcal{O}}=\Delta\, {\mathcal{O}}=(L+s+\gamma (g,s,L)) {\mathcal{O}} \, . \label {dil}
\end{equation}
We indicate with $\lambda =8\pi ^2 g^2$ the 't Hooft coupling and, for given $s,L$, we focus on the linear combination of operators with minimal anomalous dimension $\gamma (g,s,L)$. Moreover, we go to the high spin ($s\rightarrow\infty$) limit: in such a limit the anomalous dimension is proportional to $\ln s$,
\begin{equation}
 \gamma(g,s,L)=f(g)\ln
s+f_{sl}(g,L)+O\left(\frac{1}{\ln s}\right) \, , \nonumber
\end{equation}
the scaling function $f(g)$ being twice the cusp anomalous dimension of light-like Wilson loops \cite {KM}.

The main motivation of our work comes from the
AdS/CFT correspondence \cite{MWGKP}. This is a remarkable strong/weak coupling duality which, in particular, equates the spectrum of anomalous
dimensions of composite operators in ${\cal N}=4$ SYM to
the energy spectrum of states in type IIB superstring theory in AdS$_5\times $S$^5$.
Another reason of interest for twist two operators (\ref {oper}) resides in the connection between elements
of their anomalous dimension matrix and the Mellin transforms of splitting kernels of DGLAP equations.
In particular, $f_{sl}(g,2)$ coincides with the coefficient
of the $\delta (1-z)$ term, i.e. the virtual scaling function.

In order to compute anomalous dimensions in planar limit, we will use the powerful tool of integrability. Integrability was first found in
one loop planar QCD: on the one hand, scattering of Reggeised gluons was shown to be described by
spin 0 Heisenberg chain \cite {FK}; on the other hand, the spectrum of anomalous dimensions of aligned-helicity twist operators was found to coincide with the spectrum of the spin -1/2 Heisenberg chain \cite {BDM}.
In planar ${\cal N}=4$ SYM integrability exists not only at one loop \cite
{MZ}, but at all loops and in all the gauge theory sectors \cite{BS}. In brief, every composite operator can be thought of as a
state of a 'spin chain', whose Hamiltonian is the dilatation operator itself: the large size ({\it asymptotic}) spectrum is described by certain Asymptotic Bethe Ansatz (ABA) equations (the so-called Beisert-Staudacher equations, cf. \cite {BS,BES} and references therein).
Unfortunately, this works only for infinitely long operators: anomalous dimensions of operators with finite length depend not only on ABA data, but also on finite size 'wrapping' corrections \cite {WRA}. Recent progress has shown that a set of Thermodynamic Bethe Ansatz (TBA) equations \cite{TBA}
provides exact (any length at any coupling) predictions on anomalous dimensions of planar ${\cal N}=4$ SYM.
Yet, $f(g)$ and $f_{sl}(g,L)$ are wrapping-free, cf. below.

\section{From Bethe equations to linear integral equations}

We now briefly review how ABA helps in computing anomalous dimensions.
Restricting to operators $\mathcal{O}(x)$ and supposing that wrapping correction can be neglected,
anomalous dimensions $\gamma(g,s,L)$ can be expressed as
\begin{equation}
\gamma (g,s,L)=\frac {ig^2}{2}\sum _{k=1}^s \left [ \left (\frac
{1}{x^+(u_k)}\right )-\left (\frac {1}{x^-(u_k)}\right )
\right ] \, , \nonumber
\end{equation}
with $u_k$ solutions of 'Bethe' equations,
\begin{equation}
\left ( \frac {u_k+\frac
{i}{2}}{u_k-\frac {i}{2}} \right )^L  \left ( \frac
{1+\frac {g^2}{2{x_k^-}^2}}{1+\frac {g^2}{2{x_k^+}^2}} \right
)^L=\mathop{\prod^s_{j=1}}_{j\neq k}  \frac {u_k-u_j-i}{u_k-u_j+i}
 \left ( \frac {1-\frac {g^2}{2x_k^+x_j^-}}{1-\frac
{g^2}{2x_k^-x_j^+}} \right )^2  e^{2i\theta
(u_k,u_j)} \, , \label {bethe}
\end{equation}
where $x^{\pm}_k=x(u_k\pm i/2) \, ,  \
x(u)=\frac {u}{2}\left [ 1+{ \sqrt {1-\frac {2g^2}{u^2}}} \right ]$
and $\theta(u,v)$ is the dressing phase \cite {BS,BES}.
It is convenient to study (\ref {bethe}) in the high spin limit, as in \cite {BGK}.
In such a limit, indeed, wrapping corrections are reduced: perturbative computations \cite {BJL} show that
they affect $O\left (\frac{(\ln s)^2}{s^2} \right )$ terms. In addition, we showed \cite {BFR,FR} that, at high spin,
the nonlinear Bethe equations (\ref {bethe}) can be equivalently rewritten as linear integral equations, the nonlinear terms appearing at the order $O\left ( \frac{1}{s^2} \right )$.
Putting together these two facts, we deduce that both $f(g)$ and $f_{sl}(g,L)$ can be obtained from solutions of linear integral equations
for the density of Bethe roots $\sigma (u)$, descending from the ABA equations (\ref {bethe}).
In specific, when $s\rightarrow \infty$, we can split the density of Bethe roots as $\sigma (u)=\ln s \ \sigma ^{(1)}(u) + \sigma ^{(0)}(u)+ O\left (\frac{1}{\ln s}\right )$.
At leading order $\ln s$ one has the BES equation \cite {BES}
\begin{eqnarray}
&&\hat \sigma ^{(1)}(k)=F^{(1)}(k)-
\frac{g^2 k}{\sinh \frac{k}{2}} \int _{0 }^{+\infty}{dt}
e^{-\frac{t}{2}} \hat K (\sqrt{2}gk,\sqrt{2}gt)
  \hat \sigma ^{(1)}(t)
\, , \nonumber
\end{eqnarray}
with
\begin{equation}
F^{(1)}(k)=\frac{4g^2 \pi k }{\sinh \frac{k}{2}}\hat K( \sqrt{2}gk, 0) \nonumber
\end{equation}
and 'kernel'
\begin{equation}
\hat K(t,t')=\frac{2}{tt'}\left [ \sum _{n=1}^{\infty}n J_n (t) J_n (t') + 2 \sum _{k=1}^{\infty} \sum _{l=0}^{\infty} (-1) ^{k+l}c_{2k+1,2l+2}(g) J_{2k}(t) J_{2l+1}(t') \right ] \, , \nonumber
\end{equation}
which gives $f(g)=\frac{\hat \sigma ^{(1)}(0)}{\pi }$. Even if an explicit expression for $f(g)$ was not found, weak coupling \cite {ES,BES} and strong coupling \cite {CK} expansions were easily obtained.

Going further at the order $(\ln s )^0$, one has the linear integral equation \cite {BFR}
\begin{equation}
\hat \sigma ^{(0)}(k)= F^{(0)}(k)-
\frac{g^2 k}{\sinh \frac{k}{2}} \int _{0 }^{+\infty}{dt}
e^{-\frac{t}{2}} \hat K (\sqrt{2}gk,\sqrt{2}gt)
  \hat \sigma ^{(0)}(t)
\, , \label {bfr}
\end{equation}
which has the same kernel as BES equation, but different forcing term:
\begin{eqnarray}
F^{(0)}(k)&=&4g^2 \frac{\pi k }{\sinh \frac{k}{2}} \int _{0 }^{+\infty} \frac{dt}{e^{t}-1}[ \hat K (\sqrt{2}gk,\sqrt{2}gt)-\hat K (\sqrt{2}gk,0)] + \nonumber  \\
&+&  \frac {\pi L}{\sinh \frac{k}{2}}[1-J_0({\sqrt {2}}gk)]+4g^2 \gamma _E \frac{\pi k }{\sinh \frac{k}{2}}\hat K( \sqrt{2}gk, 0) + \nonumber \\
&+& g^2 (L-2) \frac{\pi k }{\sinh \frac{k}{2}}\int _{0 }^{+\infty}dt  e^{-\frac{t}{2}}\hat K (\sqrt{2}gk,\sqrt{2}gt)  \frac {1-e^{\frac {t}{2}}}{\sinh \frac {t}{2}} \, . \nonumber
\end{eqnarray}
The solution of (\ref {bfr}) gives $f_{sl}(g,L)=\frac{\hat \sigma ^{(1)}(0)}{\pi}$: one easily gets weak coupling perturbative expansion \cite {FGR4}:
\begin{eqnarray}
&& f_{sl}(g,L) = (\gamma_E - (L-2) \ln 2)f(g)+
8 (2 L-7) \zeta (3) \, \left( \frac{g}{\sqrt 2} \right)^4  -
\nonumber  \\
&& -\frac{8}{3} \left(\pi ^2 \zeta (3) (L-4)+3 (21 L-62) \zeta (5)\right) \left( \frac{g}{\sqrt 2} \right)^6 + \nonumber \\
&& +\frac{8}{15} \left(\pi ^4 \zeta (3) (3 L-13)+75 (46 L-127) \zeta (7)+5 (11 L-32) \pi ^2 \zeta (5)\right) \left( \frac{g}{\sqrt 2} \right)^8 +\ldots  \nonumber
\end{eqnarray}
The strong coupling asymptotic series \cite {FZ,FGR4} requires a slightly bigger effort:
\begin{eqnarray}
f_{sl} (g, L) &=& 2 \sqrt{2} \,  g  \Bigl[ \ln \frac{2 \sqrt 2 }{g} - 1 -
\frac{3 \, \ln 2}{2 \sqrt{2} \pi \, g} \ln \frac{2 \sqrt 2 }{g}
+ \frac{6\ln 2 -\pi +(2-L)\pi}{2 \sqrt {2} \pi g} - \nonumber \\
&-& \frac{\textrm{K}}{ 8 \pi^2 g^2} \ln \frac{2 \sqrt 2 }{g}+ \frac{4 \textrm{K}-9 (\ln 2)^2}{16\pi ^2 \,g^2}+ O(\frac{\ln g}{g^3}) \Bigr]   \label {strong} .
\end{eqnarray}
Importantly, result (\ref {strong}) agrees with string theory computations \cite {BFTT}.


\end{document}